\documentclass{aa}
\usepackage{graphics}
\newcommand{\doverd}[2]{\frac{\partial #1}{\partial #2}}
\newcommand{\doverdt}[1]{\frac{\partial #1}{\partial t}}

\newcommand{\subscr}[1]{_{\rm #1}}

\begin{document}

   \thesaurus{06     
              (03.13.4;  
               02.01.2;  
               07.16.1;  
               07.19.1;  
               08.02.3)} 
   \title{On the treatment of the Coriolis force in
        computational astrophysics}


   \author{W. Kley}


   \institute{Theoretisch-Physikalisches Institut,
              Friedrich-Schiller Universit\"at,
              Max-Wien Platz 1, D-07743 Jena, Germany\\
              email: wak@tpi.uni-jena.de
             }

   \date{Received 1998; accepted 1998}

   \maketitle

\begin{abstract}
In numerical computations of astrophysical flows one frequently has
to deal with situations which require the use of a corotating
coordinate system. Then, in addition to the centrifugal force,
the Coriolis terms have to be included in the Navier-Stokes
equations. In usual numerical treatments of solving the equations
of motion these non-inertial forces are included explicitly as source
terms. In this letter we show that this procedure, if applied
to the angular momentum equation, may lead to erroneous
results. Using the specific sample calculation of a protoplanet embedded in a
protostellar disk, we then demonstrate that only a conservative treatment   
of the Coriolis term in the angular momentum equation
yields results which are identical to those obtained by computations 
performed in the inertial frame. 

   \end{abstract}

%
%
\section{Introduction}
In numerical computations of the hydrodynamic evolution of
astrophysical flows the usage of a rotating coordinate system is
often desirable. For example, in calculating the evolution of 
an accretion disk in a cataclysmic variable system usually
a coordinate system which rotates with the orbital speed of the
secondary is used (eg. Heemskeerk, 1994). Then the secondary star
remains at a fixed location in the grid, simplifying the numerical
procedure. Also, in computations of a disk around a binary star
(Artymowicz \& Lubow, 1994) or with an embedded planet in the disk
(Kley, 1998, Bryden et al. 1998) corotating coordinates are
very useful. In particular, this simplifies grid refinement
in the vicinity of an embedded planet to resolve the detailed flow within
the planet's Roche lobe.

In standard, recent approaches to model the tidal effect
on an accretion disk in a binary system the Coriolis terms
are treated as explicit source terms in the equations of motion
(eg. Heemskerk 1994, Godon 1996, Kunze et al. 1997). Using this
procedure, the Coriolis force does not conserve total
angular momentum. We are not aware that in any of these computations
comparison runs between the inertial and corotating frame have been
performed.
As will be shown in this letter, using the sample
calculation on an embedded planet within an accretion disk,
the explicit approach yields incorrect results concerning
the overall evolution of the system.
We then proceed to demonstrate that, on the contrary,
the conservative formulation of the Coriolis term in the angular momentum 
equation leads to results which are identical to those performed in
the inertial frame.

In the following section 2 the physical setup of the calculation
is described. In section 3, the dependence
of the obtained results on the treatment of the Coriolis force
are displayed, and in section 4 we summarize.

\section{Physical Model}
In an accretion disk where the vertical
thickness $H$ is usually much smaller than the distance $r$
from the center ($H/r << 1$) it is customary to work only with
vertically averaged quantities.
Hence, the system under consideration consists of an infinitesimally thin
disk with an embedded protoplanet. We work in a rotating
coordinate system where the planet is at a fixed location.
We shall work in cylindrical coordinates ($r, \varphi, z$),
where $r$ is the radial coordinate, $\varphi$ is the azimuthal angle,
and $z$ is the vertical axis, and the rotation is around the $z$-axis, i.e
${\bf \Omega} = (0,0,\Omega)$. The origin of the
coordinate system is located at the center of mass of the
system.

In a flat disc located in the $z=0$ plane, the
velocity components are ${\bf u} = (u_r, u_{\varphi}, 0)$.
In the following we will use the symbol $v=u_r$ for the radial velocity
and $\omega = u_{\varphi}/r$ for the angular velocity of the flow,
which are measured in the corotating frame.
Then the vertically integrated equations of motion are
\begin {equation}
 \doverdt{\Sigma} + \nabla (\Sigma {\bf u} )
                =  0,  \label{Sigma}
\end{equation}
\begin {equation}
 \doverdt{(\Sigma v)} + \nabla (\Sigma v {\bf u} )
  = \Sigma r ( \omega + \Omega)^2
        - \doverd{p}{r} - \Sigma \doverd{\Phi}{r} + f\subscr{r}
      \label{u_r}
\end{equation}
\begin {equation}
 \doverd{[\Sigma r^2 \omega]}{t}
   + \nabla (\Sigma r^2 \omega {\bf u} )
         =
        - 2 \Sigma v \, \Omega \, r - \doverd{p}{\varphi} 
       - \Sigma \doverd{\Phi}{\varphi} + f\subscr{\varphi}
      \label{u_phi}
\end{equation}
Here $\Sigma$ denotes the surface density and $p$ the
vertically integrated (two-dimensional) pressure. The
gravitational potential $\Phi$ created by the protostar
with mass $M_s$ and the planet having mass $m_p$ is given by
\[
    \Phi = - \frac{G M_s}{| {\bf r} - {\bf r}_s |}
       -  \frac{G m_p}{| {\bf r} - {\bf r}_p |},
\]
where $G$ is the gravitational constant and ${\bf r}_s,
{\bf r}_p$ are the radius vectors to the star and planet, respectively.
To model the potential of the planet within the Roche lobe we introduce
a smoothing length of 1/5 of the Roche radius.
The effects of viscosity are contained in the terms $f_r$, and $f_{\varphi}$
which give the viscous force per unit area acting
in the radial and azimuthal direction. The explicit form of these terms is
given in Kley (1998).

Note, that in the angular momentum equation (\ref{u_phi}) the Coriolis forces
are included explicitly as source terms on the right hand side.
If treated numerically,
they do not conserve total angular momentum, and to demonstrate the effect
of this formulation on the physical results the following, additional
conservative formulation is used in the computations
\begin {equation}
 \doverd{[\Sigma r^2 (\omega + \Omega)]}{t}
   + \nabla [\Sigma r^2 (\omega + \Omega) {\bf u} ]
         =
        - \doverd{p}{\varphi} - \Sigma \doverd{\Phi}{\varphi}
   + f\subscr{\varphi}
      \label{u_phi2}.
\end{equation}
As will be shown, only this latter form leads to correct results in the
corotating frame.
   \begin{figure}
     \resizebox{\hsize}{!}{\includegraphics{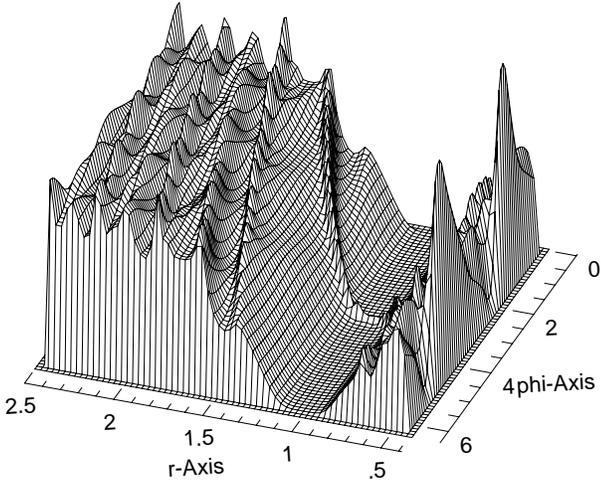}}
     \caption{3D-mountain plot of the surface density using a rotating
      coordinate system and the conservative
      treatment (Eq. \ref{u_phi2}) of the angular momentum equation}
         \label{fig1}
    \end{figure}
%
   \begin{figure}
     \resizebox{\hsize}{!}{\includegraphics{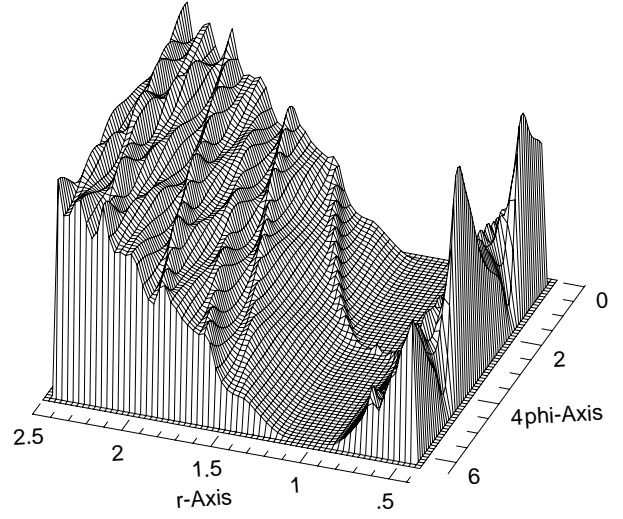}}
     \caption{3D-mountain plot of the surface density using a rotating
     coordinate system and the explicit
      treatment (Eq. \ref{u_phi}) of the angular momentum equation}
         \label{fig2}
    \end{figure}
%
   \begin{figure}
     \resizebox{\hsize}{!}{\includegraphics{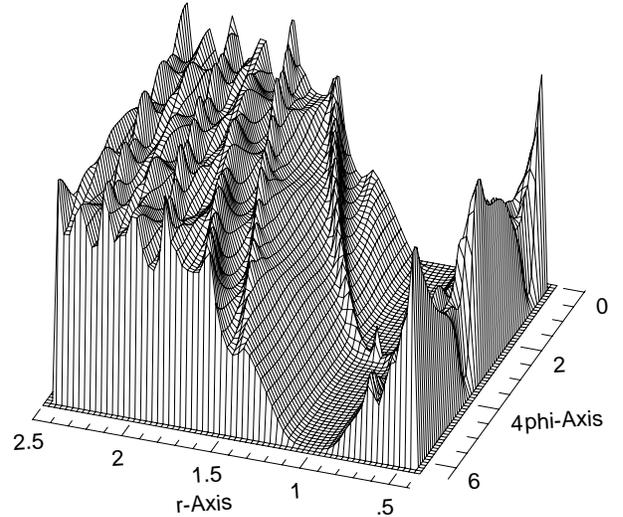}}
     \caption{3D-mountain plot of the surface density using a non-rotating
      coordinate system ($\Omega=0$)}
         \label{fig3}
    \end{figure}
An energy equation is not required since an isothermal equation
of state is used where the pressure is given by $p = \Sigma \, c\subscr{s}^2$.
The sound speed is set to a constant fraction of the
Keplerian velocity $c\subscr{s} = 0.05 v\subscr{Kep}$, which is
equivalent to a fixed relative thickness of the disk, $H/r=0.05$.
The kinematic viscosity coefficient in these test calculations is set
to a constant value $\nu = 10^{-5}$ in dimensionless units (see below),
and the bulk viscosity is set to zero.
For the mass ratio of the planet to the star we assume
$q = m_p/M_s = 10^{-3}$.

\subsection{Numerical issues, initial and boundary conditions}
To solve the equations of motion dimensionless units are used where
the unit of length is given by the distance $a$ of the planet
to the star. The unit of time is given by $t_0 = \Omega\subscr{p}^{-1}$
where $\Omega\subscr{p}$ is the orbital speed of the planet; identical
to the speed of the rotating coordinate system. The other units follow
from these. The disk is non self-gravitating, and the density
scales out of the problem. The time in the result section, and
the plots is given units of the period
$P\subscr{p}=2\pi\Omega^{-1}\subscr{p}$ of the planet. 

The detailed description of the physical model used for the present
analysis is outlined in Kley (1998) and is not repeated here. 
The main results of this letter
are obtained with a code called {\it Nirvana} (Ziegler \& Yorke, 1997),
which is a 3D-multi-grid MHD code.
It was adapted for this 2D, single grid, purely hydrodynamic
application and extended by including the viscous terms explicitly.
The code uses a spatially second order accurate, explicit method
which handles advection by a monotonic transport algorithm. Hence,
the advective parts
of the code conserve globally mass and angular momentum.
An additional run, using identical physical parameter,
has been performed with a different, independent
code {\it WK} developed by Kley (1989), which uses an implicit treatment
of viscosity.

The initial setup consists of a Keplerian disk, with a density profile
$\Sigma \propto r^{-1/2}$ which follows in case of a constant kinematic
viscosity. Additionally, an initial gap (lowering of the density)
is assumed around the planet whose size can be estimated by equating
viscous and gravitational torques (Lin \& Papaloizou, 1993).
The initial $\Sigma(r)$ profile is given for example in Fig.~4.
The radial velocity is set to zero initially. The inner and outer boundary
are closed throughout the evolution.
The computational domain has the extent $r\subscr{min}=0.4$,
$r\subscr{max}=2.5$, and covers the whole azimuthal range
$\varphi\subscr{min}=0$, $\varphi\subscr{max}=2\pi$. The grid resolution
consists of $128 \times 128$ equidistantly spaced grid points.
The physical parameter for all 4 models presented are identical, and
we state in Table 1 the main, differing computational features of the
models.
\begin{table}
\caption[]{Numerical parameter of the models}
\begin{tabular}{lll}
Model  &    Code  &  Description \\
\hline
1      &   Nirvana &  Rotating frame (conservative, Eq. \ref{u_phi2})  \\
2      &   Nirvana &  Rotating frame (explicit, Eq. \ref{u_phi})  \\
3      &   Nirvana &  Inertial frame ($\Omega =0$)  \\
4      &   WK      &  Inertial frame ($\Omega =0$)\\ 
\hline
\end{tabular}
\end{table}
\section{Numerical results}
This initially axially symmetric disk setup is evolved under the influence
of the perturbing potential of the planet, which very soon creates
non-axisymmetric features in the density distribution. Trailing spiral
arms are developing inside and outside of the planet's radial
location. For details and the influence of the variation
of the physical parameter see Kley (1998), and Bryden et al. (1998).
Here we are only interested in numerical aspects of the computations.
The total evolution time of the system covers about 200 orbital
periods of the planet.

In Figs. 1 to 3 mountain plots of the surface density distribution at t=200
are given for the different calculations, where for clearity only half
the grid points are displayed. While Figs. 1 and 2 refer
to the computation using a corotating coordinate system Fig. 3 refers
to a computation in the inertial frame where $\Omega$ has been set to zero.
In obtaining Fig. 1 the conservative treatment (\ref{u_phi2})
of the angular momentum equation has been used, while Fig.~2 was
computed using the explicit version (\ref{u_phi}).
%
   \begin{figure}
     \resizebox{\hsize}{!}{\includegraphics{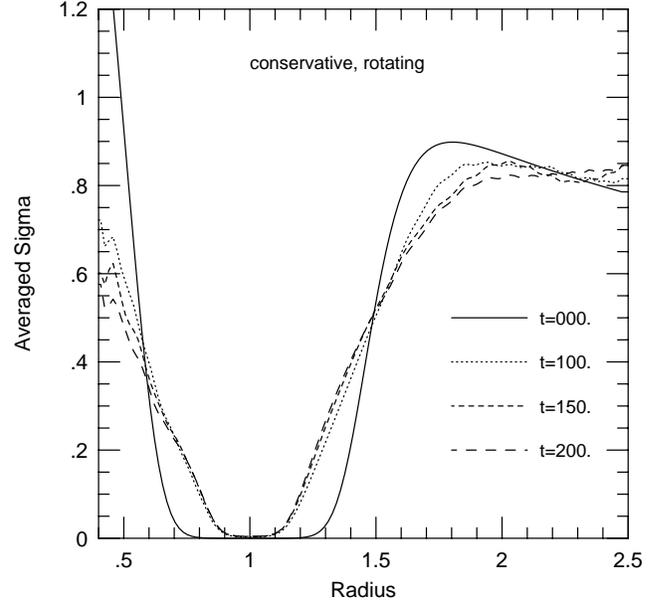}}
     \caption{The azimuthally averaged surface density for model 1.}
         \label{fig4}
    \end{figure}
%
   \begin{figure}
     \resizebox{\hsize}{!}{\includegraphics{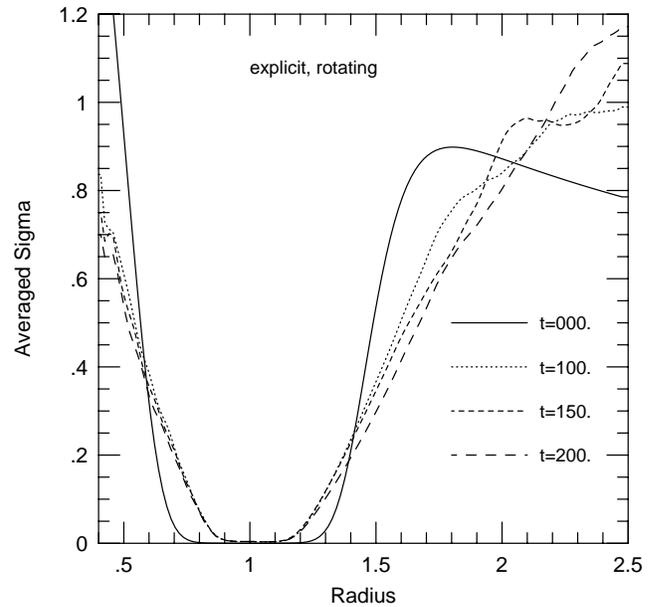}}
     \caption{The azimuthally averaged surface density for model 2.}
         \label{fig5}
    \end{figure}

Clearly, there are distinct differences in the density distribution
in the two formulations of the problem. When the explicit form
(\ref{u_phi}) of the Coriolis terms is used the obtained density profile
is much shallower radially
around the planet (located at $r=1$ and $\varphi=\pi$). More matter
seems to be pushed away from the planet to larger radii.
This indicates an additional unphysical transport of angular momentum
from inside out in the vicinity of the planet.
In contrast, the inertial frame calculation (Fig.~3) and the first,
conservative calculation (Fig.~1) agree very well with each other.
As the physical results should not depend on any assumed rotation
of the coordinates we conclude that {\it only} in the case where the
Coriolis terms are treated in a conservative fashion the correct results
are obtained.

This important finding is corroborated by looking at the time evolution
of the radial distribution of the surface density averaged over the azimuthal
angle $\varphi$ as given in Fig~4 and 5. In the conservative treatment
(Fig.~4) the averaged density distribution reaches a quasi-stationary
state which does not evolve any further.
This is to be expected once gravitational and viscous
torques are in equilibrium. In contrast, the explicit formulation
shows a continuing evolution of the density profile in the outer region
(Fig.~5), where mass appears to be accumulating near the outer
boundary.
 
A physically interesting question is related to possible mass accretion
through the gap to form more massive planets as observed in extrasolar
systems (Boss, 1996). To model such a situation, mass has been removed
continually within the Roche lobe of the planet with a half emptying time
of about 1/4 orbital periods of the planet (Kley 1998). 
In figure 6 the obtained mass accretion rate for the planets
(in dimensionless units) is plotted versus time for the four models
where the model named {\it WK} has been obtained in the inertial frame
with a different code (Kley 1989). Obviously the codes {\it Nirvana}
and {\it WK} agree very well with each other, as do the conservative
and inertial frame calculation. Only the explicit form (\ref{u_phi})
deviates again strongly and shows additionally a decline with time.
The reduced accretion onto the planet appears to be caused by artificial
angular momentum increase of the explicit model.
%
   \begin{figure}
     \resizebox{\hsize}{!}{\includegraphics{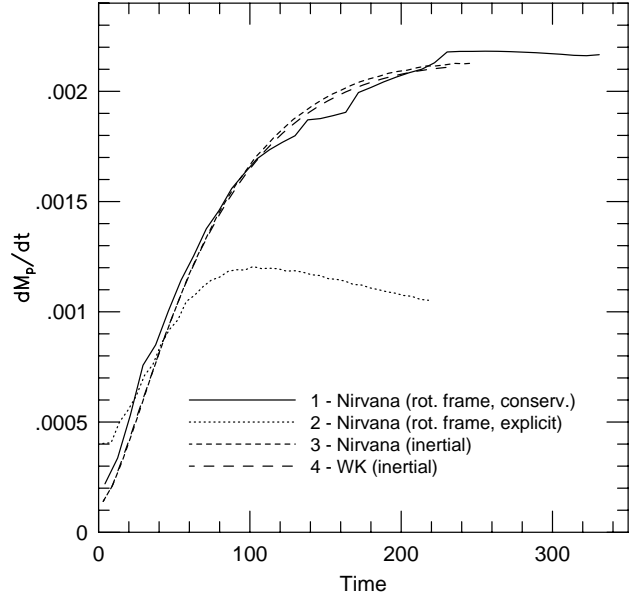}}
     \caption{The accretion rate onto the planet versus time for the different
       models.}
         \label{fig6}
    \end{figure}
\section{Local analysis}
To verify the conjecture above, we perform 
a local analysis of the finite difference
equations. Specifically, let us consider the change of
angular momentum $\Delta J$ within one gridcell during
one timestep $\Delta t$.
For simplicity, axial symmetry is assumed, and pressure and viscous
effects are not considered. The change in the $i^{th}$ radial gridcell
with volume $V_i$ using the conservative treatment is then given
purely by the advection term
\begin{equation}  \label{jcons}
     \Delta J^{cons}_i = - \left[ F_{i+1}(\omega+\Omega) a_{i+1} 
                        - F_{i}(\omega + \Omega) a_i \right] \Delta t,
\end{equation} 
with the angular momentum flux (flow per unit length and time) 
$F_i(\omega + \Omega) = r^3_i \, v_i \, (\omega_{i-1} + \Omega) \, 
\Sigma_{i-1}$, through the curcumference $a_i = 2 \pi r_i$.
To demonstrate the effect, we consider a simple upwind scheme on a staggered,
equidistant grid, where $\Sigma_i$ and $\omega_i$ are located 
at $r_{i+1/2}$ in the middle of the cell interfaces
$r_i$ and $r_{i+1}$. The radial velocities $v_i$ are located at
$r_i$, and in (\ref{jcons}) positive velocities are assumed.
For the explcicit treatment (Eq. \ref{u_phi}) we obtain
\begin{eqnarray} \label{jexpl}
     \Delta J^{expl}_i  & = & - \left[ F_{i+1}(\omega) a_{i+1} 
                        - F_{i}(\omega) a_i \right] \Delta t, \nonumber \\
          &  + & \Delta J^{Cor}_i  + r_{i+1/2}^2 \Omega \Delta M_i 
\end{eqnarray}
where 
\begin{equation}
     \Delta J^{Cor}_i = - \Sigma_i \, (v_i + v_{i+1} ) \Omega r_{i+1/2}
                           \Delta t \Delta V_i
\end{equation}
is the change in angular momentum caused by the explicit Coriolis term,
and $\Delta M_i$ is the change of mass in the gridcell.
Taking now the difference $\Delta J = \Delta J^{expl}_i - \Delta J^{cons}_i$,
and assuming 
\[
   \Sigma = \Sigma_0 \, r^s, \, \, v = v_0 r^q, \nonumber
\]
and a quasi-stationary density distribution $\Delta M_i =0$, then we find
\begin{equation}  \label{deltaj}
   \Delta J = 2 \pi \Omega \Delta t \Sigma_0 v_0 \, r^{q+s+3} \, 2x (q+s+1),
\end{equation}
where $x = \Delta r / r_{i+1/2}$.
Thus, for positive velocities which inevitably occur at the outer
edge of the gap and a positive gradient of the density (see Fig. 4 and 5),
we find that $\Delta J > 0$.  
Hence, the explicit treatment shows an artificial enhancement
of angular momentum, which tends to widen the gap created by the planet
even further (see Fig.5). We notice, that the identical relation
(\ref{deltaj}) is obtained for negative velocity ($v_0 < 0$) which means that
for a standard accretion disk with $q = -1$ and $s$ negative, the
explcicit treatment yields to an artificial increase of angular momentum
as well. Both effects play a role in the presented model
calculations of a planet embedded in an accretion disk.
\section{Conclusions}
We have performed numerical computations of a planet embedded in a disk
using different treatments of the Coriolis force in the angular
momentum equation.

Three different models, one with an explicit Coriolis force, the
second using a conservative formulation for the Coriolis term, and the third
in the inertial frame ($\Omega=0$) have been calculated.
It was seen that only the
second and the third model yield identical results, which were also
confirmed by an additional calculation using a completely independent
numerical code.

We conclude that in performing numerical models which use a rotating
reference frame, care must be taken to incorporate the Coriolis terms
correctly in the angular momentum equation. As often in numerical
computations a conservative formulation of the Coriolis force
turned out to be superior over an explicit, non-conservative treatment.
In the future numerical modelers should be aware of this potential
problem when calculating
applications such as disks in binary systems, circumbinary disks,
rotating stars, or similar problems.
\begin{acknowledgements}
The author is very grateful to Dr. Udo Ziegler for making his
code {\it Nirvana} publicly available.
Useful discussions with Dr. Steve Lubow (STScI) are gratefully
acknowledged.
This work was supported by the Max-Planck-Gesellschaft
Grant No. 02160-361-TG74.
\end{acknowledgements}

\begin{thebibliography}{}

\bibitem[1994]{arty} Artymowicz P., Lubow S.H., 1994, ApJ, 421, 651
\bibitem[1996]{boss} Boss A.P., 1996, Physics Today, 49/9, p32.
\bibitem[1998]{bryden} Bryden G., Chen X., Lin D.N.C.,
    Nelson, R.P., Papaloizou, J.C.B., 1998,
    preprint, submitted to ApJ
\bibitem[1994]{heemskerk} Heemskerk M.H.M., 1994, A\&A, 288, 807
\bibitem[1997]{godon} Godon P., 1997, ApJ, 480, 329
\bibitem[1989]{kley89} Kley, W., 1989, A\&A, 208, 98
\bibitem[1998]{kley98} Kley, W., 1998, submitted to MNRAS
\bibitem[1997]{kunze} Kunze S., Speith R., Riffert H., MNRAS, 289, 889
\bibitem[1993]{lin} Lin D.N.C., Papaloizou J.C.B., 1993, in
   {\it Protostars and Planets III}, Ed. E.H. Levy, J.I.Lunine,
    U. Arizona P., Tucson.
\bibitem[1997]{ziegler} Ziegler U., Yorke H.W., Comp. Phys. Comm., 101, 54

\end{thebibliography}
\end{document}